\documentclass[12pt,a4paper]{article}%

\usepackage{amsmath,amssymb,amsfonts}
\usepackage{caption2}
\usepackage[]{hyperref}
\usepackage[pdftex]{color,graphicx}

\numberwithin{equation}{section} 
\setcaptionmargin{1cm}

\begin{document}
\begin{titlepage}
\vskip 1.0cm
\begin{center}
{\Large \bf Minimal Flavour Violation with hierarchical squark masses} \vskip 1.0cm {\large Riccardo
Barbieri, Enrico Bertuzzo, Marco Farina, Paolo Lodone  and Dmitry Zhuridov} \\[1cm]
{\it  Scuola Normale Superiore and INFN, Piazza dei Cavalieri 7, 56126 Pisa, Italy} \vskip 1.0cm
\end{center}
\begin{abstract}
In a supersymmetric model with hierarchical squark masses we analyze a pattern of flavour symmetry breaking  centered on  the special role of the top Yukawa coupling and, by extension, of the full Yukawa couplings for the up-type quarks. For sufficiently heavy squarks of the first and second generation this
leads to effective Minimal Flavour Violation of the Flavour Changing Neutral Current amplitudes. For this to happen we determine the bounds on the masses of the heavy squarks with QCD corrections taken into account, properly including previously neglected effects.  We believe that the view presented in this paper altogether strengthens the case for hierarchical sfermions.
\end{abstract}
\end{titlepage}

\section{A pattern of flavour symmetry breaking}

The elusiveness so far
of any clear deviation from the Standard Model (SM) in Flavour Changing Neutral Current (FCNC) amplitudes poses problems to phenomenological supersymmetry, arguably even more than the lack of direct signals of any s-particle.
Since the early times the supersymmetric flavour problem has been the subject  of many investigations with several suggestions for its possible  solution.
However the remarkable progression of the flavour tests achieved in the last years has rendered the problem more acute. While deviations from the SM could be hiding just around the corner, as perhaps even hinted by recent data, the overall quantitative success of the SM in describing many measured FCNC effects calls for a reconsideration of the issue.

Broadly speaking one can group the various attempts at addressing the supersymmetric flavour problem in three categories: "degeneracy" \cite{Dimopoulos:1981zb}\cite{Barbieri:1981gn}, "alignment" \cite{Nir:1993mx} or "hierarchy"\cite{Dine:1990jd}-\cite{Giudice:2008uk}. Here we follow a specific direction centered on  the special role of the top Yukawa coupling and, by extension, of the full Yukawa couplings for the up-type quarks. We base its implementation on a definite pattern of flavour symmetry breaking, which  will result in a kind of blending of the three different approaches  just mentioned.

Our assumptions are the following:
\begin{itemize}
\item Among the squarks, only those that interact with the Higgs system via the top Yukawa coupling are significantly lighter than the others.

\item With only the up-Yukawa couplings, $Y_u$, turned on, but not the down-Yukawa couplings, $Y_d$, there is no flavour transition between the different families.

\end{itemize}
The first assumption is in line with the "hierarchical" picture. On one side the tight constraints on the flavour structure of the first two generations of squarks, $\tilde{q}_{1,2}$,  by kaon physics get relaxed by taking $\tilde{q}_{1,2}$ sufficiently heavy. On  the other side the naturalness upper bounds on all the squarks that do no feel the top Yukawa couplings, i.e. again $\tilde{q}_{1,2}$ and the right-handed sbottom, are much looser than for all the other s-particles\cite{Dimopoulos:1995mi}.
The second assumption corresponds to the "alignment" between $Y_u$ and the
squared mass matrices of the left-handed doublet squarks, $m^2_{\tilde{Q}}$,  and of the right-handed singlet squarks of charge 2/3, $m^2_{\tilde{u}}$. This alignment can result from a suitable pattern of flavour symmetry breaking.

In fact, in the $Y_d = 0$ limit, the largest flavour symmetry consistent with the above hypotheses
is
\begin{equation}
U(1)_{\tilde{B}_1}\times U(1)_{\tilde{B}_2}\times U(1)_{\tilde{B}_3}\times U(3)_{d_R},
\label{U3}
\end{equation}
where $\tilde{B}_i$ acts as baryon number but only on the supermultiplets $\hat{Q}_i$ and $\hat{u}_{R_i}$ of the i-th generation, respectively the left-handed doublets and the charge-2/3 right-handed singlets, whereas $U(3)_{d_R}$ acts on the three right-handed supermultiplets of charge 1/3.
We are going to analyze in detail the consequences of this flavour symmetry, assumed to be broken down to overall baryon number  by the small  $Y_d$ couplings only. Throughout this paper we take $\tan{\beta}$, as usually defined in supersymmetric models, below about 10.

 Smaller symmetries that are interesting to consider as well are
\begin{equation}
U(1)_{\tilde{B}_1}\times U(1)_{\tilde{B}_2}\times U(1)_{\tilde{B}_3}\times U(1)_{d_{R_3}}\times U(2)_{d_R}
\label{U12}
\end{equation}
and
\begin{equation}
\Pi_{i=1}^3 U(1)_{\tilde{B}_i} \times U(1)_{d_{R_i}},
\label{U123}
\end{equation}
always broken  by the (supersymmetric) down-Yukawa couplings only. Needless to say, when $Y_d$ is switched on, $U(3)_{d_R}$  still implies approximate degeneracy of all the right-handed down squarks, whereas only the  first two generations are approximately degenerate in the $U(2)_{d_R}$ case.
These symmetries can be compared, for $Y_u= Y_d =0$, to
\begin{equation}
U(3)_{Q}\times U(3)_{u_R}\times U(3)_{d_R},
\label{U33}
\end{equation}
that leads, under suitable further hypotheses, to Minimal Flavour Violation (MFV)\cite{D'Ambrosio:2002ex} of the FCNC amplitudes.

\section{Flavour changing transitions}

Let us  analyze the consequences of (\ref{U3}). In the physical basis for the charge 2/3 quarks, where we work from now on, the squared mass matrices $m^2_{\tilde{Q}}, m^2_{\tilde{u}}$ and the A-terms for the charge 2/3 squarks are flavour diagonal, up to possible corrections controlled by $Y_d$. We promote $Y_d$ to a non-dynamical spurion field transforming  under (\ref{U3}) in such a way that the down-quark Yukawa couplings, $H_d  \bar{Q}_L Y_d d_R$, are formally invariant. In this case the symmetry-breaking
 corrections to the diagonal mass matrices are at least quadratic in $Y_d$ and  can be safely neglected.  This is  in view of the non degeneracy of $m^2_{\tilde{Q}}, m^2_{\tilde{u}}$ and, in particular, of the large separation between the third and the first two  eigenvalues.
 As a consequence, the mass matrices of the up-type squarks, left or right, and of the left down-type squarks are diagonal.
This is unlike the case of the right handed down-squark squared mass matrix, for which
\begin{equation}
m^2_{\tilde{d}_R} = m^2 ({\bf 1} + a Y_d^+ Y_d),
\label{m2d}
\end{equation}
where $m$ sets the scale of the $\tilde{d}_R$-squark masses,  {\bf 1} is the $3\times 3$ unit matrix and $a$ is an unknown numerical coefficient.

Other than (\ref{m2d}), the only other mass matrix that needs to be diagonalized is the $d$-quark mass matrix,
\begin{equation}
\mathcal{L}_m (d-quarks) = (v \cos{\beta}) \bar{d}_L Y_d d_R + h.c.
\label{md}
\end{equation}
where $v \approx 175$ GeV. With $Y_d$ expressed in terms of two unitary matrices and its diagonal form, $Y_d = V y_d U^+$, it is now immediate to go to the full physical basis for all the matter fields. The matrix $U$ can be transformed away both from (\ref{m2d}) and (\ref{md}) by a simultaneous unitary rotation of the $\hat{d}$ supermultiplets without affecting any interaction term.
On the contrary, by $d_L \rightarrow V d_L$,  the matrix $V$ enters as the only flavour-changing matrix in the  interaction terms expressed in the physical basis for all matter fields
\begin{eqnarray}
\mathcal{L}_{FC} &=& \frac{g}{\sqrt{2}}\, (\overline{u_L} \gamma^\mu V\, d_L) \, W^+_\mu - g\, \tilde{u}^*_L V \,\overline{\tilde{W}^- }\, d_L \,+ \frac{g}{\sqrt{2}}\, \tilde{d}_L^* V \, \overline{\tilde{W}^3}\, d_L \nonumber \\
&& -\sqrt{2} \frac{g'}{6} \tilde{d}^*_L\, V \,\overline{\tilde{B}}\, d_L -\sqrt{2}\, g_3 \, \tilde{d}_L^* \, \lambda^b\, V\, \overline{\tilde{g}^b }\,d_L \nonumber \\
&& + \tilde{{u}}_R^* \, y_u \, V \, \overline{\tilde{H}_{u}^- }\, d_L+\overline{{u}_R} \, y_u\, V \,d_L \, H_u^+ + h.c.,\nonumber \\
\label{LFC}
\end{eqnarray}
where  terms proportional to $Y_d$ have been neglected. As seen from the first term on the right-hand side of (\ref{LFC}), $V$ is the Cabibbo-Kobayashi-Maskawa matrix. In deriving (\ref{LFC}) we have  neglected the mixing between left and right squarks induced by the A-terms. Their introduction would not change the fact that $V$ is the only  flavour-changing matrix. Furthermore,  mild conditions on their size make them only relevant in the $\tilde{t}_L- \tilde{t}_R$ mixing, which can be straightforwardly introduced in the analysis.
$\mathcal{L}_{FC}$ in (\ref{LFC}) coincides with the one that would be obtained from (\ref{U33}) and $Y_u, Y_d$  were promoted to spurions that keep the supersymmetrized SM Yukawa couplings invariant\footnote{The degeneracy of all the squark masses as $Y_u, Y_d = 0$ requires in fact some further assumptions on the relative size of the various corrections  induced by switching on $Y_u$ and $Y_d$\cite{Kagan:2009bn}.}. The squark spectrum is, in the two cases, largely different.

The interactions in (\ref{LFC}) inserted in suitable box diagrams give rise to the following general structure of the $\Delta F = 2$ effective Lagrangian
\begin{equation}
\mathcal{L}^{\Delta F =2} =
\Sigma_{\alpha \neq \beta} \Sigma_{j,k} \xi_k^{\alpha\beta} \xi_j^{\alpha\beta} f_{j,k} (\bar{d}_{L\alpha}\gamma_\mu d_{L \beta})^2 + h.c.,~\alpha, \beta =d, s, b;~j, k = 1,2,3,
\label{FCNC2}
\end{equation}
where $\xi_j^{\alpha  \beta} = V_{j \alpha} V_{j \beta}^*$ (here $j = u,c,t$) and
$f_{j, k} = f_{k,j}$ are functions of the masses of the j-th, k-th generations of up or down squarks, of the charged Higgs boson and of the various gaugino, higgsinos.

Similarly from penguin-type diagrams one obtains
\begin{equation}
\mathcal{L}^{\Delta F = 1} =
\Sigma_s \Sigma_{\alpha \neq \beta} \Sigma_k \xi_k^{\alpha\beta}  f_k^{(s)} Q^{\alpha\beta}_{(s)} + h.c.,
\label{FCNC1}
\end{equation}
where $s$ extends over all the the effective operators $Q^{\alpha\beta}_{(s)}$ relevant to  the processes with different final states in $\Delta F=1$ FCNC transitions and the functions $f_k^{(s)}$  depend on the masses of the  k-th generations of up or down squarks, other than on the masses of the various gaugino, higgsinos and of the charged Higgs boson.

To be precise, both (\ref{FCNC2}) and (\ref{FCNC1}) only include the extra contributions from the SM ones, due to the standard charged current interaction in (\ref{LFC}), which however, in the down sector, have exactly the same structure. The only difference is in the form of the functions $f_{j,k}$ and $f_k^{(s)}$, which, in the SM case, depend on the W mass and on the masses of the up-type quarks of the j-th, k-th generations.

\section{Effective Minimal Flavour Violation}

Using $\Sigma_i \xi_i^{\alpha\beta}  = 0$ for any $\alpha\neq\beta$, it is useful to reorganize the various terms in  (\ref{FCNC2}) as
\begin{equation}
\mathcal{L}^{\Delta F =2}  = \mathcal{L}^{\Delta F =2}_{33} +
\mathcal{L}^{\Delta F =2}_{12} + \mathcal{L}^{\Delta F =2}_{12,3}
\label{F2dec}
\end{equation}
where
\begin{equation}
\mathcal{L}^{\Delta F =2}_{33} =
\Sigma_{\alpha \neq \beta} ( \xi_3^{\alpha\beta})^2 (f_{3,3} - 2 f_{3,1} + f_{1,1}) (\bar{d}_{L\alpha}\gamma_\mu d_{L \beta})^2 + h.c.,
\label{FCNC33}
\end{equation}
\begin{equation}
\mathcal{L}^{\Delta F =2}_{12} =
\Sigma_{\alpha \neq \beta} ( \xi_2^{\alpha\beta})^2 (f_{2,2} - 2 f_{2,1} + f_{1,1}) (\bar{d}_{L\alpha}\gamma_\mu d_{L \beta})^2 + h.c.,
\label{FCNC12}
\end{equation}
\begin{equation}
\mathcal{L}^{\Delta F =2}_{12,3} =
\Sigma_{\alpha \neq \beta} 2 \xi_2^{\alpha\beta} \xi_3^{\alpha\beta} (f_{3,2} -  f_{3,1} + f_{1,1} - f_{1,2}) (\bar{d}_{L\alpha}\gamma_\mu d_{L \beta})^2 + h.c.
\label{FCNC123}
\end{equation}
Similarly, for the $\Delta F=1$ case,
\begin{equation}
\mathcal{L}^{\Delta F =1}  = \mathcal{L}^{\Delta F =1}_{31} +
\mathcal{L}^{\Delta F =1}_{21}
\label{F1dec}
\end{equation}
where
\begin{equation}
\mathcal{L}^{\Delta F = 1}_{31} =
\Sigma_s \Sigma_{\alpha \neq \beta} \xi_3^{\alpha\beta} ( f_3^{(s)}- f_1^{(s)}) Q^{\alpha\beta}_{(s)} + h.c.,
\label{FCNC131}
\end{equation}
\begin{equation}
\mathcal{L}^{\Delta F = 1}_{21} =
\Sigma_s \Sigma_{\alpha \neq \beta} \xi_2^{\alpha\beta} ( f_2^{(s)}- f_1^{(s)}) Q^{\alpha\beta}_{(s)} + h.c.
\label{FCNC121}
\end{equation}
In the SM, where, as said, these expressions also apply, the high degeneracy of the up and charm quarks relative to the W mass makes such that only the first terms on the right-hand-sides of (\ref{F2dec}) and (\ref{F1dec}) (for brevity "the top-quark exchanges") are relevant, or dominate over the others, in every FCNC process, with the exception of the "real part" of the $\Delta S=2$ transition, where $\mathcal{L}^{\Delta F =2}_{12}$ is important. Therefore, in any extension of the SM where
the extra FCNC effects are described by (\ref{FCNC2}) and (\ref{FCNC1})  and, furthermore, the first term on the right-hand-side of (\ref{F2dec}) and (\ref{F1dec}) dominates over the others, all FCNC amplitudes have the forms
\begin{equation}
\mathcal{A}^{\Delta F=2}_{\alpha\beta}|_{MFV} = \mathcal{A}^{\Delta F=2}_{\alpha\beta}|_{SM} (1 + \epsilon^{\Delta F=2})
\end{equation}
\begin{equation}
\mathcal{A}^{\Delta F=1, s}_{\alpha\beta}|_{MFV} = \mathcal{A}^{\Delta F=1, s}_{\alpha\beta}|_{SM} (1 + \epsilon^{\Delta F=1, s})
\end{equation}
with $ \epsilon^{\Delta F=2}$ and $\epsilon^{\Delta F=1, s}$ real and universal, i.e. not dependent on $\alpha$ and $\beta$.
This is called effective Minimal Flavour Violation. For clarity of the exposition we are neglecting here flavour blind CP phases\cite{Dugan:1984qf}.

As already mentioned  a supersymmetric extension of the SM with a maximal flavour symmetry (\ref{U33}) only broken by $Y_u$ and $Y_d$  leads under reasonable assumptions to effective MFV.  In this case  the extra terms in (\ref{F2dec}) and (\ref{F1dec})  are suppressed by the high degeneracy of the squarks of the first two generations relative to their mean masses.
In the case of hierarchical squark masses considered here, based upon (\ref{U3}),  the extra terms in  (\ref{F2dec}) and (\ref{F1dec}) 
will in general only be suppressed by the heaviness of the first and second generation of squarks.
From the dependence upon $\xi_i^{\alpha\beta}$ of $
\mathcal{L}^{\Delta F =2}_{12}, \mathcal{L}^{\Delta F =2}_{12,3}$ and $\mathcal{L}^{\Delta F = 1}_{21}$
and the consideration of the experimental constraints on the various FCNC amplitudes,
it is  seen that the dominant effects to be taken under control to obtain effective MFV are:
\begin{itemize}
\item From $\mathcal{L}^{\Delta F =2}_{12}$ the contribution to the "real part" of $\Delta S =2$;
\item From $\mathcal{L}^{\Delta F =2}_{12,3}$ the contribution to the "imaginary part" of $\Delta S =2$.
\end{itemize}
Furthermore, once these constraints are satisfied, all  possible deviations from MFV in other FCNC channels are negligibly small.

\section{Inclusion of QCD corrections}
\label{QCDc}

The precise knowledge of the mixing angles allows a neat determination of the bounds to be satisfied by the heavy squark masses to obtain effective MFV. To this end resummed QCD corrections must also be taken into account.
Since there is no other  $\Delta S = 2$ operator involving the left-handed fields $d_L, s_L$  other than
\begin{equation}
Q_1 = (\overline{d}^\alpha \gamma^\mu P_L {s}^\alpha) \, (\overline{d}^\beta \gamma_\mu P_L {s}^\beta)
\label{Q1}
\end{equation}
(with colour indices made explicit), one would think that the QCD corrections consist in a simple rescaling of the well known anomalous dimension of $Q_1$. This is true for the Lagrangian  $\mathcal{L}^{\Delta F =2}_{33}$,
 with the corresponding box diagrams in leading order  only sensitive to "low" momenta (of the order of the masses of the lighter squarks and of the gluino) and for $ \mathcal{L}^{\Delta F =2}_{12}$, generated by box diagrams sensitive only to "high" momenta (of the order of the masses of the heavier squarks).
 This is however not the case of $ \mathcal{L}^{\Delta F =2}_{12,3}$ with the corresponding box diagrams sensitive, already  in leading order, to all momenta between the low and the high scale\footnote{
The calculation of the QCD corrections to the $\Delta S =2$ effective Lagrangian has in fact already been considered in \cite{Bagger:1997gg}\cite{Agashe:1998zz}\cite{Contino:1998nw} in the context of a hierarchical spectrum, but the special problem presented by $ \mathcal{L}^{\Delta F =2}_{12,3}$ was missed.}.

Let us deal for simplicity only with the gluino box diagrams which give, in most of the parameter space, the dominant contribution.
The new ingredient that is required to deal with the heavy-light exchange in $\mathcal{L}^{\Delta F =2}_{12,3}$ is the mixing between the $\Delta S = 2$ operator (\ref{Q1})
and the $\Delta S = 1$ operators with two gluino external legs, for which a possible basis is
$$
\begin{array}{lll}
Q_1^g &=&\delta^{ab}\delta_{\beta\alpha}(\overline{d}^\beta P_R \widetilde{g}^b)(\overline{\widetilde{g}^a}P_L s^\alpha )\\
Q_2^g &=& d^{bac} t^c_{\beta\alpha}(\overline{d}^\beta P_R \widetilde{g}^b)(\overline{\widetilde{g}^a}P_L s^\alpha )\\
Q_3^g &=&i f^{bac} t^c_{\beta\alpha}(\overline{d}^\beta P_R \widetilde{g}^b)(\overline{\widetilde{g}^a}P_L s^\alpha )\;.\\
\end{array}
$$
The appropriate effective Lagrangian  to work with is
\begin{equation}
\mathcal{L}^{eff} = C_1 Q_1 + \Sigma_i C_{i}^g Q^g_i
\end{equation}
It is convenient to define the scale-dependent 4-component vector
\begin{equation}
{\bf C}^T = (C_1, \hat{C}_g^T);~ \hat{C}_g^T = (C^g_1, C^g_2, C^g_3)
\end{equation}
satisfying an appropriate initial condition at $\mu = m_h$, a mean heavy squark mass, and a Renormalization Group Equation (RGE)
\begin{equation}
\frac{d {\bf C}}{d \log{\mu}} = { \Gamma}^T {\bf C}.
\end{equation}
The $4\times 4$ matrix of anomalous dimensions, $\Gamma$, receives contributions both from standard gluon exchanges as from flavour-changing light-squark exchanges. Its explicit expression for a generic $SU(N)$ of colour is
$$
\Gamma=\frac{\alpha_s}{2\pi}\left(
\begin{array}{cc}
\gamma _1& \xi^{ds}_3 \hat{\gamma}_{1g}\\
\xi^{ds}_3 \hat{\gamma}_{g1}^T &  \hat{\gamma}_{gg}\\
\end{array}
\right),
$$
where $\gamma_1 = 3\frac{N-1}{N}$ is the standard anomalous dimension of $Q_1$ and:
\begin{equation}
\hat{\gamma}_{g1}  =  \left(\frac{N^2-1}{4 N}, \frac{N^2-4}{8N}\frac{N-1}{N},  \frac{N-1}{8}\right)
\label{eq:gammas}
\end{equation}
\begin{equation}
\hat{\gamma}_{gg}=
\left(
\begin{array}{ccc}
\frac{n_\ell}{4} & 0 & -6 \\
0 & -\frac{3N}{2} + \frac{n_\ell}{4} & -\frac{3N}{2} + \frac{6}{N} \\
-3 & -\frac{3N}{2} & -\frac{3}{2}N + \frac{n_\ell}{4}
\end{array}
\right),
\end{equation}
where $n_\ell$ is the number of light squarks ($\tilde{t}_L \, , \, \tilde{t}_R \, , \, \tilde{b}_L$, i.e. $n_\ell=3$ in our context).

We are interested in the expression for $C_1(m_{\ell})$ at the light scale, with $m_{\ell} \approx m_{\tilde{g}} \approx m_{\tilde{Q}_3}$, up to first order in $\xi_3^{d s}$ (which makes $\hat{\gamma}_{1g}$ irrelevant).
To this end one has first to evolve the $\hat{C}_g$ to the scale $\mu$, which is readily done by diagonalizing the $3\times 3$ matrix $\hat{\gamma}_{gg}$, via $\hat{\gamma}_{gg}^T = A \hat{\gamma}_{gg}^D A^{-1}$.
In terms of $A$ and of the diagonal matrix $\hat{\gamma}_{gg}^D$, one has
\begin{equation}
\hat{C}_g (\mu) = A \left( \frac{\alpha_s(\mu)}{\alpha_s(m_h)}\right)^{\hat{\gamma}_{gg}^{D} /b_0} A^{-1} \hat{C}_g (m_h),
\label{eq:solDF1}
\end{equation}
where $b_0/2\pi$ is the first coefficient of the beta-function for $\alpha_s$.

The RGE for $C_1$ now has the form
\begin{equation}
\frac{dC_1}{d\log \mu}= \frac{\alpha_s}{2\pi}\left( \gamma_1 C_1 + \xi_3^{ds} \hat{\gamma}_{g1} \hat{C}_g \right),
\label{eq:RGEDF2}
\end{equation}
where the last term on the right-hand-side, upon insertion of (\ref{eq:solDF1}),  is a known function of $\mu$. By standard techniques one has therefore
\begin{equation}
C_1 (m_{\ell}) = \left( \frac{\alpha_s(m_{\ell})}{\alpha_s(m_h)}\right)^{\gamma_1/b_0}C_1 (m_h)+ \xi_3^{ds}  \hat{\gamma}_{g1}  A B_D A^{-1}\hat{C}_g (m_h)\;,
\label{eq:solDF2}
\end{equation}
with the matrix elements of the diagonal matrix $B_D$ given by
\begin{equation}
(B_D)_{kk}= \frac{1}{\gamma_{k}-\gamma_1} \left[ \left( \frac{\alpha_s(m_{\ell})}{\alpha_s(m_h)}\right)^{\gamma_k/b_0}-\left(  \frac{\alpha_s(m_{\ell})}{\alpha_s(m_h)}\right)^{\gamma_1/b_0}\right]\; ,~\gamma_k = (\hat{\gamma}^D_{gg})_{kk}.
\label{eq:Gamma}
\end{equation}
The first term on the right-hand-side of (\ref{eq:solDF2}) corresponds to the standard rescaling of $Q_1$, whereas the second term, proportional to $\xi_3^{ds}$ is the QCD corrected contribution appearing at lowest order in $\mathcal{L}^{\Delta F =2}_{12,3}$. Indeed by expanding the second term in $\alpha_s$ one recovers the lowest order coefficient of the form $\xi_2^{ds} \xi_3^{ds} (\alpha_s^2/m_h^2) \log{m_h/m_{\ell}}$.

\section{Lower bounds on the masses of the heavy squarks}

We are in the position to determine the lower bound that have to be satisfied by the heavy squark masses in order to give rise to effective MFV in the sense discussed in Section 3. For simplicity we take
\begin{equation}
m_{\tilde{u}_{R1}} \approx
m_{\tilde{Q}_{L1}} \equiv m_1,~~
m_{\tilde{u}_{R2}} \approx
m_{\tilde{Q}_{L2}} \equiv m_2
\end{equation}
The full expressions of the functions $f_{j,k}$ entering in (\ref{FCNC2}) can be found in the literature\cite{Bertolini:1990if}\cite{Gabbiani:1996hi}. To allow an analytic control of the final results we describe the two limiting cases:
\begin{itemize}
\item Quasi Degenerate: $\delta \equiv 2 (m_1^2 - m_2^2)/ (m_1^2 + m_2^2)$ sufficiently small that an expansion in $\delta$ can be made (and $m_1^2 + m_2^2 \equiv 2 m_h^2$)
\item Non Degenerate: $m_1 >> m_2$ (or, equivalently, viceversa).
\end{itemize}
We also take all the masses of the light particles comparable to each other and to the mass $m_{\ell}$, i.e.
\begin{equation}
m_{\tilde{g}} \approx m_{\tilde{t}_L} \approx m_{\tilde{t}_R} \approx m_{\tilde{b}_L} \approx m_{\chi^\pm} \approx m_{\chi^0} \approx m_{\ell}.
\end{equation}
From the lowest order box diagrams one obtains
\begin{equation}
\mathcal{L}^{\Delta S=2}_{12} = S_{12} Q_1 + h.c.
\label{L}
\end{equation}
where, for the two limiting cases (the indices $s, d$ on $\xi_{2,3}$ omitted in this Section for brevity):
\begin{equation}
S_{12}^{QD} = \xi_2^2 \frac{\delta^2}{m_h^2}\left[ \alpha_s^2 \frac{11}{108}+ \frac{\alpha_w^2}{72}\left(12 \mathcal{R}^2+ 8\frac{\alpha_s}{\alpha_w}\mathcal{R} + 3\right)\right],
\end{equation}
\begin{equation}
S_{12}^{ND} = \frac{\xi_2^2}{m_2^2}\left[\frac{11}{36}\alpha_s^2 + \frac{\alpha_w^2}{24} \left(3+ 8\frac{\alpha_s}{\alpha_w} \mathcal{R}+12\mathcal{R}^2\right) \right],~
\mathcal{R} = \frac{1}{\cos^2{\theta_W}}
\left( \frac{1}{4} -\frac{2}{9}\sin^2{\theta_W}\right)
\end{equation}
Similarly, for the heavy-light exchange at lowest order one has
\begin{equation}
\mathcal{L}^{\Delta S=2}_{12,3} = S_{12,3} Q_1 + h.c.
\end{equation}
where
\begin{equation}
S_{12,3}^{QD} =
\xi_2 \xi_3 \frac{\delta}{m_h^2}\left[ \alpha_s^2 \left(-\frac{35}{18}+ \frac{11}{18}\log{\frac{m_h^2}{m_{\ell}^2}}\right) +\alpha_w^2\left(\frac{1}{4} +  \left(\frac{1}{4}+ \frac{2}{3}\frac{\alpha_s}{\alpha_w} \mathcal{R}+\mathcal{R}^2\right)\left( -4+\log{\frac{m_h^2}{m_{\ell}^2}}\right)\right)\right]
\end{equation}
\begin{equation}
S_{12,3}^{ND} =
 \xi_2 \xi_3 \frac{1}{m_2^2}\left[ \alpha_s^2 \left(-\frac{37}{36}+ \frac{11}{18}\log{\frac{m_2^2}{m_{\ell}^2}}\right) +\alpha_w^2\left(\frac{1}{4} +  \left(\frac{1}{4}+ \frac{2}{3}\frac{\alpha_s}{\alpha_w} \mathcal{R}+\mathcal{R}^2\right)\left( -\frac{5}{2}+\log{\frac{m_2^2}{m_{\ell}^2}}\right)\right)\right]
 \label{NDS}
\end{equation}
Both in the QD as in the ND cases we are neglecting terms vanishing as $m_{\ell}^2/m_h^2$.

From the above equations, using the results of the  previous Section, the $\alpha_s^2$ terms in $\mathcal{L}^{\Delta S=2}_{12} + \mathcal{L}^{\Delta S=2}_{12,3}$ can be corrected to include the resummed higher-order QCD effects by computing $C_1(m_{\ell})$.   The relevant  initial conditions at the heavy scale, to be used in (\ref{eq:solDF2}), are:
\begin{itemize}
\item For the Quasi Degenerate case:
\begin{equation}
C_1 =\frac{\alpha_s^2}{m_h^2}\left( \xi_2^2 \frac{11}{108}\delta^2 - \xi_2\xi_3 \frac{35}{18}\delta\right) ;~ \hat{C}_g^T = - 4\pi \alpha_s\xi_2 \frac{\delta}{m_h^2} \left( \frac{1}{3}, 1, 1 \right)
\end{equation}

\item For the Non Degenerate case:
\begin{equation}
C_1 = \frac{\alpha_s^2}{m_2^2}\left( \xi_2^2 \frac{11}{36}-\xi_2\xi_3 \frac{37}{36}\right) ;~ \hat{C}_g^T = -4\pi \alpha_s\xi_2 \frac{1}{m_2^2} \left( \frac{1}{3}, 1, 1 \right)
\end{equation}
\end{itemize}
$C_1(m_{\ell})$ can  then be evolved down to the GeV scale in a standard way, properly accounting for the different thresholds one encounters in the beta-function coefficient.
\begin{center}
 \begin{figure}[tb]
 \centering
\includegraphics[width=0.46\textwidth]{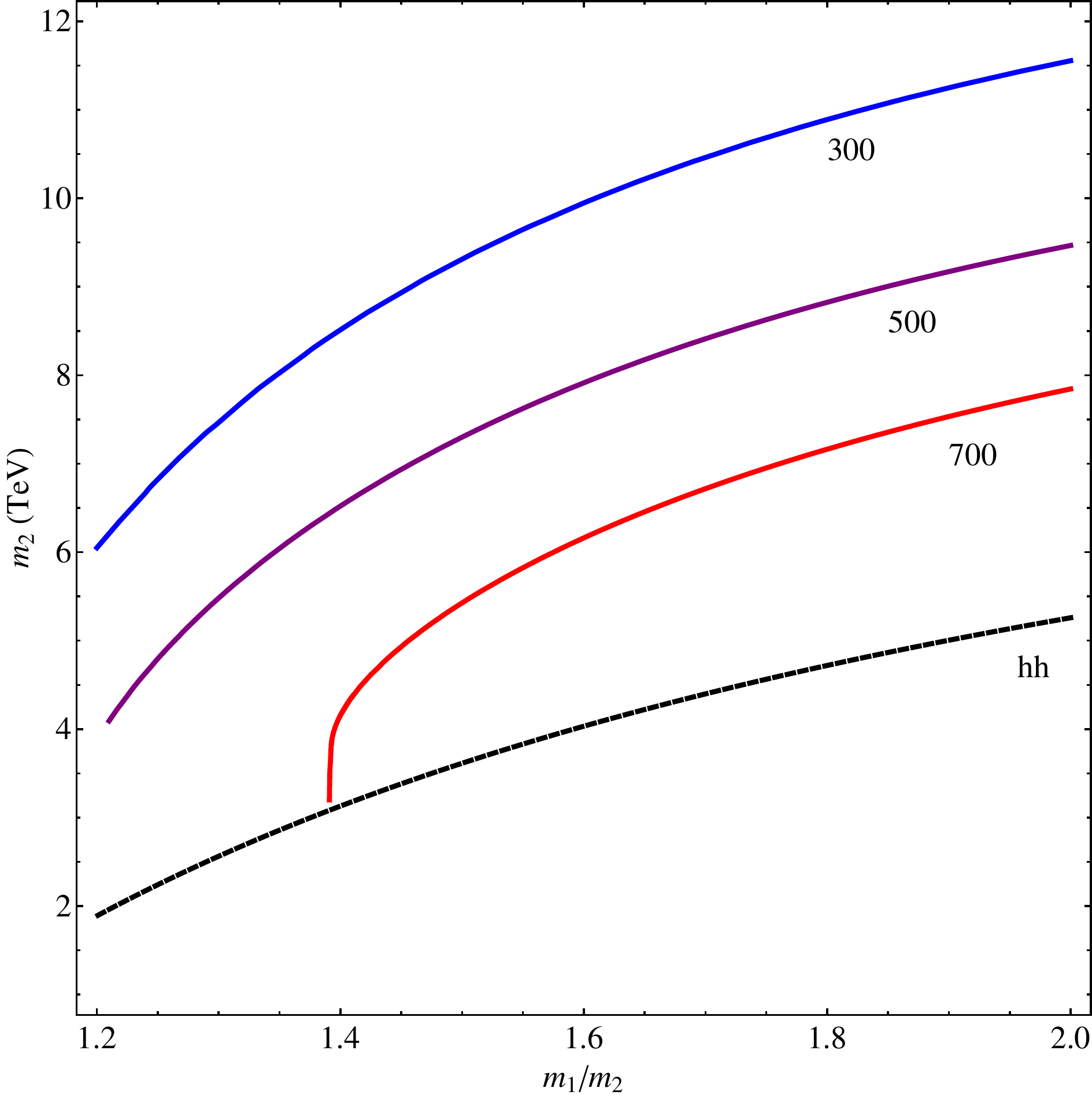}
   \caption{Lower bounds on  $m_2$ as a function of the ratio $r=m_1/m_2$ to obtain effective MFV. For a given light mass, $m_{\ell}=300,~500,~700$ GeV, the allowed region is above the corresponding line, from $\mathcal{L}^{\Delta S=2}_{12,3}$, and in any case above the "hh" line, from $\mathcal{L}^{\Delta S=2}_{12}$, which is $m_{\ell}$ independent.}
   \label{fg:bound_light-heavy}
 \end{figure}
 \end{center}
To determine finally the lower limits on the heavy squark masses that give rise to effective MFV in the FCNC amplitudes we use the following bounds, quoted in \cite{Isidori:2010kg} and referred to the  parametrization
\begin{equation}
\mathcal{L}^{\Delta S=2} =\pm \left(\frac{1}{\Lambda_{Re}^2} + \frac{i}{\Lambda_{Im}^2}\right) Q_1 + h.c.,
\label{Lambda_def}
\end{equation}
with the standard definition of the phases of the quarks $s$ and $d$:
\begin{itemize}
\item $\Lambda_{Re} > 9.8\cdot 10^2$ TeV, relevant to $\mathcal{L}^{\Delta S=2}_{12}$, which depends on the two heavy masses, $m_1$ an $m_2$;
\item $\Lambda_{Im}> 1.6\cdot 10^4$ TeV, relevant to $\mathcal{L}^{\Delta S=2}_{12,3}$, which depends on the two heavy masses, $m_1$ an $m_2$ and on the light mass $m_{\ell}$.
\end{itemize}
The lower limits implied by these bounds on  $m_2$ are shown in Fig. \ref{fg:bound_light-heavy} as function of the ratio, $r=m_1/m_2$, from $r=1.2$ to $r=2$. Given our hypotheses we would not be able to defend a too near degeneracy of the two heavy masses. For values of $r$ higher than 2 all the curves rapidly flatten  out since the heavier mass decouples. The bound from $\mathcal{L}^{\Delta S=2}_{12,3}$ is shown for three different values of $m_{\ell}$. For any given value of $m_{\ell}$ what determines the bound on $m_2$ is the strongest between the one derived from the heavy - heavy exchange (from $\mathcal{L}^{\Delta S=2}_{12}$ and denoted "hh" in the figure) and that arising from $\mathcal{L}^{\Delta S=2}_{12,3}$.
The near equality between $\xi_2$ and $\xi_1$ implies that the bounds shown in this figure would be almost identical if the ratio between $m_1$ and $m_2$, the masses of the first two generations of squarks, were reversed.
As seen from Fig.\ref{fg:bound_light-heavy},  in most cases the bound is dominated by the limit  on $\mathcal{L}^{\Delta S=2}_{12,3}$ from $\Lambda_{Im}$  in (\ref{Lambda_def}), which makes the QCD corrections computed in the previous Section particularly relevant\footnote{The precision of this bound could  be further improved with the inclusion of QCD corrections also to terms in (\ref{L}-\ref{NDS}) containing the electroweak couplings. This requires a straightforward extension of the calculation described in the previous Section.}.

\section{Less restrictive symmetry breaking patterns}

As mentioned in Section 1, it is of interest to consider also the case in which the symmetry (\ref{U3}) is lowered to 
(\ref{U12}) or (\ref{U123}). In this case it is no longer true that the unitary matrix $U$ that diagonalizes $Y_d$ on the right can be transformed away without affecting the various interactions. On the contrary, in the physical basis for the various particles, the flavour changing Lagrangian in (\ref{LFC}) receives the extra contribution
\begin{equation}
\Delta\mathcal{L}_{FC} = -\sqrt{2} \frac{g'}{3} \tilde{d}_R^*\, U \,\overline{\tilde{B}}\,d_R +\sqrt{2}\, g_3 \, \tilde{d}_R^* \, \lambda^b\, U\,\overline{\tilde{g}^b}\, d_R + h.c.
\label{DLFC}
\end{equation}
with a serious loss of predictive power since $U$ is unknown. We nevertheless present an estimate of the dominant effects, always in the form of lower bounds on the masses of the heavy squarks in order to maintain effective MFV.
To this end we define $\eta_j^{\alpha\beta} = U_{j\alpha} U^*_{j\beta}$ and, at least as a normalization, we consider
\begin{equation}
 \eta_j^{\alpha\beta} = \xi_j^{\alpha\beta} e^{i\phi^{\alpha\beta}_j}
\end{equation}
where $\phi^{\alpha\beta}_j$ are arbitrary phases. Furthermore we do not assume any special degeneracy among the squark mass parameters that respect (\ref{U12}) or (\ref{U123}).

Under these assumptions the largely dominant contribution to the FCNC amplitudes arises  again in the CP violating $\Delta S=2$ channel. Due to the much larger hadronic matrix elements of the left right operators
\begin{equation}
Q_{4,5} = (\bar{d}_R s_L) (\bar{d}_L s_R)
\end{equation}
(with two possible contractions of the colour indices), by similar arguments to the ones used in Section 3 one easily sees that the most important effects are:
\begin{itemize}
\item For the symmetry (\ref{U12})
\begin{equation}
\Delta \mathcal{L}^{\Delta S = 2, LR}_{123} =
\xi_2\eta_3 (g_{23} - g_{21} - g_{13} + g_{11}) Q_{4,5}
\end{equation}

\item For the symmetry (\ref{U123})
\begin{equation}
\Delta \mathcal{L}^{\Delta S = 2, LR}_{12} =
\xi_2\eta_2 (g_{22} - g_{21} - g_{12} + g_{11}) Q_{4,5}
\end{equation}

\end{itemize}
with the functions $g_{ij}$, not symmetric in $i, j$, dependent upon  left and right down squark masses. For all the heavy down squark masses of typical size $m_h$, it is
\begin{equation}
\Delta \mathcal{L}^{\Delta S = 2, LR}_{(123, 12)}\approx
(\xi_2\eta_3, \xi_2\eta_2) \frac{\alpha_s^2}{m^2_h} Q_{4,5}
\label{DLS2}
\end{equation}
up to dimensionless coefficients of order unity, sensitive to the ratios of the squark masses. If one uses (\ref{DLS2})
for $Q_4$, in a standard notation, and one scales down its coefficient by its diagonal anomalous dimension, ignoring mixing with $Q_5$, one obtains the lower bounds:
\begin{itemize}
\item For the symmetry (\ref{U12})
\begin{equation}
m_h \gtrsim    450~TeV     \left(\left|\frac{\eta_3}{\xi_3}\right| \sin{\phi_3}\right)^{1/2}
\end{equation}

\item For the symmetry (\ref{U123})
\begin{equation}
m_h \gtrsim    10^4~TeV     \left(\left|\frac{\eta_2}{\xi_2}\right| \sin{\phi_2}\right)^{1/2}
\end{equation}
\end{itemize}
Once again, without deviations from the assumptions made on the various  parameters (or special relations among them), these bounds imply effective  MFV in all other FCNC amplitudes to high precision.


\section{Summary and conclusions}

The idea that the squarks with only a small coupling to the Higgs system, i.e. the first two generations and perhaps the right handed sbottom as well, be significantly heavier than the other ones, the two stops and the left handed sbottom, has received a lot of attention for different reasons. A particular motivation  has been seen in the context of the supersymmetric flavour problem, as of the supersymmetric CP problem. It is also well known, however, that solving these problems by purely raising the masses of the first two generations of squarks without further specific assumptions requires values for these masses far beyond any reasonable naturalness limit, as in fact explicitly shown in the previous Section. The  progress of the last decade in testing the flavour structure of the SM strengthens the motivations to reconsider this subject.

This same experimental progress has brought the focus on the so called MFV. While it is clear that MFV is far from being a theory of flavour, 
it may nevertheless contain an element of physical reality in as much as it rests on a postulated pattern of flavour symmetry breaking. In the context of supersymmetry, the example of $U(3)_{Q}\times U(3)_{u_R}\times U(3)_{d_R}$ only broken in a suitable way by $Y_u$ and $Y_d$ certainly offers a possible way to address the flavour problem. 
At the same time it is clear that $U(3)_{Q}\times U(3)_{u_R}\times U(3)_{d_R}$ is not compatible with hierarchical sfermions.
In this paper we have shown that effective MFV can also be made compatible with  hierarchical sfermions as long as the relevant flavour symmetry is $U(1)_{\tilde{B}_1}\times U(1)_{\tilde{B}_2}\times U(1)_{\tilde{B}_3}\times U(3)_{d_R}$, only suitably broken by the small $Y_d$ couplings. 
This is the case if the heavy squark masses satisfy a definite lower bound, that we have quantified in a precise way,
 as summarized in Fig.\ref{fg:bound_light-heavy}. This    bound is dominated  by the limit  on $\mathcal{L}^{\Delta S=2}_{12,3}$ from $\Lambda_{Im}$  in (\ref{Lambda_def}), which makes the QCD corrections computed in Section \ref{QCDc} particularly relevant.

We are not discussing any dynamical theory capable of producing the phenomenological pattern that we are advocating. Such pattern requires a coupling between the sources of flavour breaking and of supersymmetry breaking, which may at first seem difficult to realize. We think however that 
too little is known about these matters to consider this as a serious obstacle.
At the time being it suffices to notice that the pattern we propose is centered on  the special role of the top Yukawa coupling and, by extension, of the full Yukawa couplings for the up-type quarks, with the down Yukawa coupling matrix as a small perturbation. We believe that the view presented in this paper altogether strengthens the case for hierarchical sfermions.

The study of signals of flavour with supersymmetry and MFV, with or without flavour blind CP phases, is beyond the scope of this work and has in fact already received lot of attention since long time \cite{Duncan:1983iq}-\cite{Altmannshofer:2007cs}.  
One point that we find useful to emphasize in this context, however, is the importance of the value of $\tan{\beta}$.
Most if not all the significant deviations from the SM still possible in supersymmetry with MFV, which  occur in $\Delta F =1$ and in CP-violating $\Delta F =0$ amplitudes, 
are related to the possibility that $\tan{\beta}$ acquires a large value, like $\tan{\beta}\approx m_t/m_b$, which has not been our concern here\cite{Altmannshofer:2009ne}. On the other hand some of us have recently put forward a view that relates the  hierarchical sfermions with a heavier Higgs boson than in the normal Minimal Supersymmetric Standard Model\cite{Barbieri:2010pd}. This is possible for example in "$\lambda$-SUSY"\cite{Harnik:2003rs}\cite{Barbieri:2006bg}, i.e. in a Next to Minimal Supersymmetric Standard Model with a largish value of the Yukawa coupling $\lambda$ between the extra singlet $S$ and the Higgs doublets $H_{1,2}$. In  turn this requires indeed a moderate value of $\tan{\beta}$, below $3\div 4$. Whether or not it will ever be possible to detect  relatively small flavour signals in such case with hierarchical fermions is hard to tell. Definitely more likely is the earlier appearance of characteristic signals in direct production of supersymmetric particles\cite{Barbieri:2010pd}.

\section*{{Acknowledgments}}

{We thank  Gino Isidori for many useful comments and Raffaele T. D'Agnolo for his contribution in the early stages of this work. 
This work is supported in part by the European Programme ``Unification in the LHC Era",  contract PITN-GA-2009-237920 (UNILHC) and by 
MIUR under contract 2006022501. }

\end{document}